\documentclass[12pt,preprint]{aastex}

\usepackage{timesfonts}

\usepackage{psfig}

\begin{document}

\title{The birth place of gamma-ray bursts: abundance gradients and
constraints on progenitors }

\author{Enrico Ramirez-Ruiz$^1$, Davide Lazzati$^1$ and Andrew
W. Blain$^{1,2}$}
\affil{$^{1}$Institute of Astronomy, Madingley Road, Cambridge,
CB3 OHA, England; enrico,lazzati@ast.cam.ac.uk \\
$^{2}$Department of Astronomy, California
Institute of Technology, Pasadena, CA 91125, USA; awb@astro.caltech.edu\\}

\begin{abstract}
The physics of gamma-ray bursts (GRBs) and their
offsets from the centers of their host galaxies  are used to investigate
the evolutionary state of their progenitors, motivated by the popular 
idea that GRBs are linked with the
cataclysmic collapse of massive stars. We suggest that
GRB progenitors in the inner and outer regions of hosts may be intrinsically
different: outer bursts appear to have systematically 
greater isotropic equivalent energies (or narrower jets). 
This may provide an
interesting clue to the nature of GRBs, and could reflect
a relation between metallicity 
and the evolution of GRB progenitors. If true,
then this offset--isotropic luminosity correlation is 
a strong argument for a collapsar origin of
long-duration GRBs.

\end{abstract}

\section{Introduction}
One can understand the dynamics of GRB  afterglows 
simply, independent of uncertainties about their progenitors, 
using the relativistic generalization of the theory of
supernova remnants. The basic model for GRB 
hydrodynamics is of a relativistic blast wave that expands into the
surrounding interstellar medium (ISM; 
M\'esz\'aros \& Rees 1997), decelerates on contact
with the ambient matter, and  leads to a
predictable radiative spectrum with a characteristic power-law
decline. The study of GRB afterglows has provided confirmation of
relativistic source expansion (Piran 1999; M\'esz\'aros 2001). 
The energy source of the
fireball is assumed to be a cataclysmic event,
either a compact stellar merger (Lattimer \& Schramm 1976; Eichler
et~al.\ 1989) or the collapse of a massive star (Woosley 1993;
Paczynski 1998; MacFadyen \& Woosley 1999, hereafter MW99).

Evidence is accumulating that GRBs are intimately linked
with the deaths of massive stars.  For the long-burst
afterglows localized so far, the host galaxies show signs of the ongoing
star formation activity necessary for the presence of young, massive
progenitor stars (Kulkarni$\;$ et al. 1998; 
Fruchter et al. 1999; Berger et al. 2001).  The physical
properties of the afterglows,  their locations in host galaxies
(Bloom, Kulkarni \& Djorgovski 2001b), iron line
features (Piro et al.~2000; Amati et al.~2000), 
and evidence for supernova
components several weeks after three bursts (GRB980326, Bloom et
al. 1999a; GRB970228, Reichart 1999; GRB 000911, Lazzati et al. 2001)
strongly support the idea that the most common GRBs are linked to the
collapse of massive stars.\\ 

The 
circumburst medium provides a natural laboratory for studying GRBs. 
Stars that readily shed their envelopes
have short jet-crossing times and are more likely to
produce a GRB. Stars with less radiative mass loss retain a hydrogen
envelope, in which a poorly
collimated jet is likely to lose energy and fail to breaking out of the star
(MW99). 
Finding useful diagnostics for the progenitors 
is simplified if the metallicity of and physical conditions in the local ISM
influences the evolution
of the progenitor. GRBs occur close to the birth sites
of their short-lived progenitors, and so their evolution
is likely to be affected only by local properties of the host galaxy.
Here, we show that bursts
located closer to the center of their parent galaxies  have  smaller
isotropic equivalent energies (or broader jets), and so
progenitors in inner and outer galactic locations  may be intrinsically
different.  We suggest that this
could be  the outcome of  
abundance gradients in the host galaxy. 
We assume  $H_0 = 65\,\, {\rm km} \,
{\rm s}^{-1} \, {\rm Mpc}^{-1}$, $\Omega_{\rm matter}=0.3$, and
$\Omega_{\Lambda}=0.7$.

\section{The offset of GRBs from their  parental galaxies}

Important information may be gained by studying the location of GRBs
and host galaxies (Bloom et al. 1999b; 2001c). This
approach was successful for studying SN progenitors
even before detailed models of  light-curves were available (e.g. Reaves
1953). Unfortunately this kind of observation is impossible for GRB
host galaxies, as current
instruments can only resolve
circumburst environments with sizes of tens of parsecs at low
redshifts $z \approx 0.1$, and so a physical understanding of the
local GRB environment was thought to have to wait for {\it NGST}. 
Nonetheless, Bloom et
al. (2001c)  show that the distribution of the offsets of a 
small subset of GRBs
with accurate positions from 
the centers of their host galaxies is an important probe of 
their progenitors. In our analysis, we consider all 16 bursts from 
Bloom et al. (2001c) with measured angular offsets, inferred physical 
projections, secure redshifts and K-corrected, isotropic equivalent burst 
energy
estimates $E_{\rm iso}$ 
(Bloom, Frail \& Sari 2001a), and also the recently imaged GRB
010222 (see Table 1).         
We searched for correlations between the normalized offsets
of the bursts from the brightest component of their host system
($r_0$, offset/half-light radius) and 
their inferred physical properties, namely the external particle
density and the isotropic equivalent energy of the jet. The discovery
of a correlation 
could provide constraints on progenitor models. In particular, we
investigated the dependence of $E_{\rm iso}$ on
galactic location $r_0$. We found
a marginally significant correlation, 
with the innermost bursts being least energetic (Fig. 1); a
similar result is obtain using the measured physical projections, $R_0$
(see Table 1). \\

We fit a model for the correlation between $E_{\rm iso}$ and $r_0$
that takes into account 
the relative influence of each datum and its errors. 
We constructed the individual
probability distribution $p_i(x,y)dxdy$ of the true offset
at some distance $x$ and $y$ from the measured offset location
$(x_0,y_0)$, assuming that the errors in $x$ and $y$ are
uncorrelated. The probability distribution should
appear Gaussian when the offset is large,
but clearly departs from a Gaussian form for small offsets, for which
the ratio between the offset and its error is close to
unity (see Fig. 10 of Bloom et al. 2001c). 
The distribution of bursts in the log $r_0$ -- log
$E_{\rm iso}$ plane can be well modeled by a normal
distribution about a straight line (Fig. 1). To evaluate this 
correlation we created
synthetic sets of observed data from the probability
distributions of the measured values of both $r_0$ and $E_{\rm iso}$,
assuming that 
the uncertainties in $E_{\rm iso}$ are
Gaussian distributed. We then determined
model parameters and their uncertainties by fitting $10^3$ synthetic sets
of data from Monte-Carlo realizations. We find that 
the correlation extends for $\approx$ 3 orders of magnitude in $E_{\rm
iso}$, and has a positive slope, $m$, with a probability $P(m<0)
\approx 3.2 \sigma$.   The best-fit model is shown in Fig. 1 as a solid
line: $r_0 \propto E_{\rm iso}^{\approx 0.3_{-0.1}^{+0.2}}$. This positive
correlation could result from abundance
gradients in the host galaxies and so some intrinsic scatter
is expected (see Section 3).

There are some necessary limitations to our 
approach: we used only a subset of moderate-redshift bursts with
$R<28$ optical host galaxies, and well-localized
afterglows at optical and 
radio wavelengths. Both high-redshift ($z>3$) and
heavily dust-enshrouded host galaxies could be missing. More
importantly,  
dimmer bursts in the outskirts of galaxies may be missed owing to the
average decrease in density of the ISM, $n$, which
will lead to a systematic reduction in the afterglow brightness. 
This effect may be very important, but 
the afterglow flux depends on density as 
$F_\nu \propto n^{1/2}$, and so large variations in $n$ are
required to have noticeable effects: densities in the 0.1 to
50 cm$^{-3}$ range can accommodate the broadband emission of most
afterglows (Panaitescu \& Kumar 2001). Moreover, the densities derived
for these bursts do not correlate with their location in the host galaxy. This
could be due in part to the certainly diverse fractal structure
expected in the ISM. 
On the other hand, it is possible that the afterglows of
bursts close to the galactic center are more likely to suffer 
dust extinction than those in the outer parts: this 
effect may open up a scatter in the correlation, as a
greater fraction of the luminosity function becomes visible near
the edges of the galaxy. It
is also important to note that both 
the assignment of a certain observed galaxy as the host of a
GRB and the position of its center are uncertain. However, Bloom et
al. (2001c) find that the probability of a chance association
is small $<10^{-4}$; in most cases,
the apparent host has only one bright component which is assigned
as its center.

Recently, it was suggested (Frail et al. 2001; Panaitescu \& Kumar
2001; Piran et al. 2001) that the total energy 
output of GRBs is constant, and that a distribution of jet
opening angles causes the apparent dispersion in $E_{\rm iso}$. 
This analysis assumes that the breaks observed in many GRB
afterglow light-curves are due to a geometrical beam effect (Rhoads
1997) and not to either a transition to
non-relativistic expansion (Huang, Dai \& Lu 2000) or an environmental
effect such as a sharp density gradient (Chevalier \& Li 2000;
Ramirez-Ruiz et al. 2001). If the energy output of GRBs is fixed, 
then our correlation may imply a link between jet opening angle and
burst location.

\section{Abundance gradients and the physics
of GRB progenitors}

An exciting recent development in observational cosmology 
has been the extension of studies of 
abundances from the local Universe to high redshifts. The dependence
of metallicity on environment appears to be stronger than  on the
redshift of formation: 
galaxies selected using the same techniques have
metallicities rather independent of redshift, and 
old stars are not necessarily metal-poor
(Pettini 2001). Chemical abundances within different galaxies
depend strongly  on luminosity and 
environment (e.g. Vila-Costas \& Edmunds 1992; Zaritsky, Kennicut \&
Huchra 1994; Henry \& Worthey 1999; Pettini 2001). From the
center to the outermost 10 kpc, metallicity typically decreases by a
factor of ten. A comparable 
change in metallicity only occurs over a range of a factor of
a thousand in luminosity (see Fig. 5 of Pettini 2001). This is
a much greater range of luminosity than displayed by moderate-redshift GRB host
galaxies, which 
usually have magnitudes $R \approx$
25 (Table\,1). These host galaxies are UV-bright
(Trentham, Ramirez-Ruiz \& Blain  2001), and so may exhibit comparable  
abundance
gradients  to their local counterparts. Drawing inferences about GRB hosts
from local galaxies is difficult, however, since both merging and secular
evolution are likely to be important and will complicate a direct comparison. 
Nonetheless, a direct association between 
abundance gradients in GRB hosts and in local
galaxies could be responsible for the correlation
presented in Fig.\,1. 

Low-metallicity stars, which are likely to be
more prominent in the outskirts of the galaxy,  are smaller and have less
mass loss than their metal-rich counterparts. Both
properties inhibit the loss of angular momentum (MW99), and so 
low-metallicity stars are likely to be
rotating rapidly. Equatorial accretion may thus be delayed 
and a 
funnel may be produced along the rotation axis. For higher rotational
velocities this evacuated region will be more
collimated, reducing  the jet opening angle. Furthermore, for a given
mass-loss rate, the lower the metallicity, the higher both the WR
stellar mass, and the mass threshold for the removal of the hydrogen
envelope by stellar winds. These effects all increase the mass of the helium
core and favor black hole formation (MW99, Ramirez-Ruiz et al. 2001). 
If there are 
abundance gradients in the hosts, then the likely
metallicity  dependence of both black-hole formation and 
rotation suggests that GRBs in outer galactic
locations may be more energetic (greater helium core mass) or less
collimated (faster rotation) than those close to the galactic center. In the
local Universe, regular galaxies are found to have steeper abundance
gradients than those with  complex morphologies (Zaritsky et
al. 1994).  Indeed, it is reassuring that the most regular GRB host 
galaxies (shown as open circles in Fig.\,1)
firmly support the 
trend between $E_{\rm iso}$ and $r_0$. 
Note that the scatter in Fig. 1 can be due
to the dependence of metallicity on luminosity. A more detailed
analysis of the underlying reasons for the correlation requires
a large and unbiased sample of GRBs hosts, and knowledge of  
both the underlying GRB and afterglow
luminosity functions.

\section{Consequences of a dependence of GRB properties on local 
metallicity} 

What are the potential effects of a significant dependence of 
GRB luminosity, as detected by unextinguished $\gamma$-ray 
photons, on their location in the host galaxy, which could reflect 
the metallicity of their progenitors? 
The most significant is a potential offset between the true star-formation 
rate and that traced by GRB. If GRBs in outlying, 
low-metallicity environments and in low-mass galaxies are more luminous, then 
they are likely to 
be overrepresented in GRB samples, and especially in the bright BATSE 
catalog, as compared with 
those in high-metallicity environments. 

The radial dependence of metallicity $Z$ in low-redshift spiral 
(Zaritsky et al.\ 1994) and elliptical galaxies (Henry \& Worthey 1999), 
is $Z \propto \exp{(-1.9R / \overline{R})}$, while the dependence of
metallicity at fixed radius on enclosed mass $M_{\rm
enc}$ in
spiral galaxies  derived from Fig.\,4 of Henry \& Worthey is 
$Z \propto M_{\rm enc}^{\simeq -0.5}$. These functions both depend 
strongly on radius. Therefore, 
it is likely that local environmental effects will overcome global 
enrichment effects (Pettini et al.\ 2001), but that there will be a gradual 
increase in the typical luminosity of GRBs with increasing redshift
(see Lloyd-Ronning, Fryer \& Ramirez-Ruiz 2001).  

Low-mass galaxies are likely to have statistically lower metallicities 
and thus contain more luminous GRBs 
than high-mass galaxies. 
As galaxy mass is expected to build up monotonically through mergers, 
then it is possible that the highest-redshift GRBs could be systematically 
more luminous due to the lower mass of their hosts, 
perhaps by a factor of 
2--3 at $z \simeq 3$. This effect is likely to be
more significant than, but in the same direction as, 
the global increase in metallicity with cosmic time.

The most luminous GRBs 
of all could be associated with metal-free Population-III stars; 
however, 
their very high redshifts would  make examples difficult 
to find even in the {\it Swift} catalog of hundreds of bursts. 

Star-formation activity is likely to be enhanced in merging galaxies. 
In major mergers of gas-rich spiral galaxies, 
this enhancement takes place primarily in the
inner kpc, as bar instabilities drive gas into the core 
(Mihos \& Hernquist 1994). Metallicity gradients in the gas are likely to 
be smoothed out, 
both by mixing prior to star formation, and by 
SN enrichment during the burst of activity. 
GRB luminosities could thus be suppressed 
in such well-mixed 
galaxies, making GRBs more difficult to detect in these
most luminous objects, in which a significant fraction of all
high-redshift star formation is likely to have
occurred. Shocks in tidal tails associated with merging galaxies are 
also likely to precipitate the formation of high-mass stars, yet as
tidal tails are likely to consist of relatively low-metallicity
gas, it is perhaps these less intense sites of star-formation at large
distances from galactic radii that are 
more likely to yield detectable GRBs.

For star formation taking place in both merging and quiescent high-redshift 
galaxies, there should thus be a bias in favor of detecting GRBs
at a greater projected distance from the host galaxy than the mean 
radius of the star-formation activity. Hence, based on the correlation 
shown in Fig.\,1, we predict that the radial distribution of a large sample
of GRBs around their host galaxies should be considerably more 
extended than the signatures of star-formation regions within the host, 
such as blue colors, location of H$\alpha$ emission, intense radio 
emission etc. This might have the unfortunate 
consequence of making GRBs more difficult to use  as clean markers of 
high-redshift star-formation activity. Detailed observations of
the astrophysics of individual GRB host galaxies may be essential 
before  a large sample of bursts can be interpreted. More optimistically, 
the astrophysics of star formation in high-redshift 
galaxies could perhaps be studied using the intrinsic properties of a
well-selected population of GRB with deep, resolved host galaxy images.

If confirmed in detailed studies, a metallicity selection 
effect for GRBs may be able to explain the differences between the 
star-formation rate infered from observations of galaxies (Steidel et
al.\ 1999; Blain et al.\ 1999), which tend
not to increase with redshift beyond $z \simeq 2$, 
and the rate inferred from GRB counts assuming a variability--luminosity 
relation (Fenimore \& Ramirez-Ruiz 2001; Lloyd-Ronning et al. 2001), 
which continues to increase to the highest redshifts. This increase
may reflect a bias to detecting  
high-redshift GRBs in more numerous, low-mass, low-metallicity high-redshift 
galaxies. 

Another test of the effect could be provided by a comparison of the 
luminosity function of GRB host galaxies with that of the total galaxy 
luminosity function over the same redshift range. If there is a
bias towards the discovery of GRBs in low-metallicity regions, then the 
GRB host galaxy luminosity function should be biased to low
luminosities by an increasing amount as redshift increases. \\

\section{Conclusion}
We report a correlation between the isotropic equivalent energy of GRBs and
their position offset from their host galaxies. This is possibly due
to a dependence of the end point of massive stellar
evolution on metallicity. 
If confirmed in further host observations, this correlation
will both complicate interpretation of GRBs as tracers of cosmic star
formation, and potentially allow a new probe of the astrophysics in
high-redshift galaxies.

\begin{acknowledgements}
We thank G. Denicol\'o, N. Lloyd-Ronning, M. Pettini, M. J. Rees,
C. Tout and the referee for useful comments and suggestions.  ERR
acknowledges support from CONACYT, SEP and the ORS foundation. AWB
thanks the Raymond \& Beverly Sackler Foundation for financial support
at the IoA.

\end{acknowledgements}

\clearpage

\begin{table}
\caption{Properties of GRBs and host galaxies with known redshifts}
{\vskip 0.75mm}
{$$\vbox{
\halign {\hfil #\hfil && \quad \hfil #\hfil \cr
\noalign{\hrule \medskip}
Burst & $z$ & $E_{\rm iso}^{\rm a}$  ($10^{51} {\rm erg}$) &
$R_0^{\rm b}$ (kpc) & $r_0^{\rm b}$ & Host R mag$^{\rm c}$ &\cr
\noalign{\smallskip \hrule \smallskip}
970228 & 0.695 & 22.4 $\pm$ 2.50 & 3.266 $\pm$ 0.259 & 1.37 $\pm$
0.25 & 24.6 &\cr
970508 & 0.835 & 6.33 $\pm$ 0.82 & 0.09 $\pm$ 0.09
& 0.03 $\pm$ 0.03  & 25.8  &\cr
970828 & 0.958 & 249 $\pm$ 21.7 & 4.05 $\pm$ 4.33 & 1.63 $\pm$
1.80 & 24.5 & \cr
971214 & 3.418 & 185 $\pm$ 51.6 & 1.1 $\pm$
0.56 & 0.43 $\pm$ 0.23 & 25.6 & \cr 
980613 & 1.100 & 5.67 $\pm$ 1.0 & 0.78 $\pm$ 0.67
& 1.37 $\pm$ 1.57  & 26.1 &\cr 
980703 & 0.966 & 121 $\pm$ 16.0 & 0.96 $\pm$
0.54 & 0.62 $\pm$ 0.39 & 22.8 &\cr 
990123 & 1.600 & 3280 $\pm$ 512 & 6.11 $\pm$
0.03 & 2.09 $\pm$ 0.63 & 23.9 & \cr 
990506 & 1.300 & 874 $\pm$ 144 & 2.680 $\pm$ 4.144 & 2.47 $\pm$ 3.96
&25.0 & \cr   
990510 & 1.619 & 168 $\pm$ 27.1 & 0.60 $\pm$
0.08 & 0.44 $\pm$ 0.15 & 28.5 &\cr 
990705 & 0.850 & 270 $\pm$ 20.2 & 7.17 $\pm$
0.78 & 0.79 $\pm$ 0.06 & 22.8 &\cr 
990712 & 0.433 & 5.27 $\pm$ 0.67 & 0.30 $\pm$ 0.49 & 0.20 $\pm$ 0.32 &
24.4& \cr  
991208 & 0.706 & 147 $\pm$ 19.8 & 1.51 $\pm$
0.75 & 0.60 $\pm$ 0.35 & 24.4 & \cr 
991216 & 1.020 & 564 $\pm$ 79.3 & 3.11 $\pm$
0.28 & 1.27 $\pm$ 0.40 & 24.9 &\cr 
000301C & 2.033 & 46.4 $\pm$ 6.2 & 0.62  $\pm$ 0.06 & 0.44 $\pm$ 0.14
& 27.8& \cr 
000418 & 1.119 & 297 $\pm$ 99.0 & 0.20 $\pm$ 0.56 & 0.07 $\pm$ 0.19 &
23.9& \cr  
010222 & 1.476 & 712 $\pm$ 83.0 & 1.23 $\pm$ 1.30 & 0.79 $\pm$ 0.83
&$>$24.0& \cr  
\noalign{\smallskip \hrule}
\noalign{\smallskip}\cr}}$$}
\begin{list}{}{}
\item[
$^{\mathrm{ }}$]$^{\rm a}$The isotropic, K-corrected, equivalent
energies (20-2000 keV; Bloom et al. 2001a).\\
$^{\rm b}$The projected physical offset $R_0$ and the host
normalized offset (offset/half-light radius) $r_0$ are  taken from
Bloom et al. (2001c). The values for GRB 010222 are derived from 
Fruchter et al. (2001). The
associated uncertainties in the observed offsets do not necessarily
represent the   1$\sigma$ confidence region of the true offset since
the probability distribution is not Gaussian (Bloom et al. 2001c). \\ 
$^{\rm c}$Djorgovski et al.~(2001) and Trentham et al. (2001). 
\end{list}
\end{table}

\clearpage

\centerline{\psfig{file=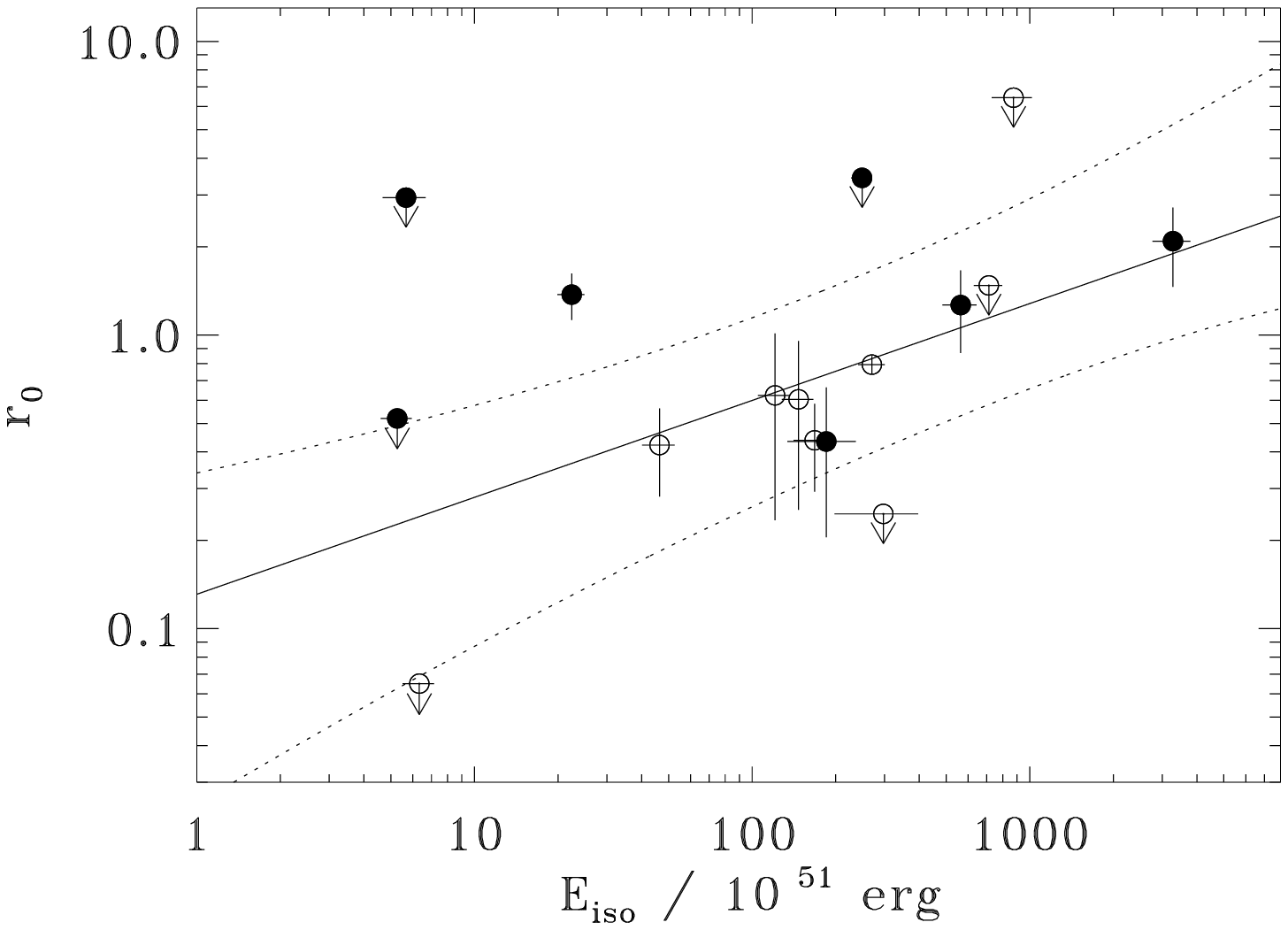,width=0.8\textwidth}} 
\figcaption{{The projected observed offset of GRBs from
their parental galaxy as a function of the burst isotropic equivalent
energy. The center of the assigned host is determined as the centroid
of the brightest component of the host system. The fractional
isophotal offsets are the observed offsets $R_0$ normalized by the
host half-light radius. Solid and dotted lines mark the center and
1$\sigma$ widths of the best-fit model distributions parameters. The
filled circles are bursts 
that occur in the most irregular, possibly merging galaxies, while the
empty circles are bursts with more regular hosts. There is a tentative
trend: the inner most bursts seem to be less energetic (similar trend
is obtained when the projected physical offsets in
kpc are plotted against the equivalent isotropic energy).} 
\label{fig1}}

\end{document}